\documentclass{iaus}
\usepackage{graphicx}
\renewcommand\vec[1]{\boldsymbol{#1}}

\title[Thermal convection in rectilinear shearing flow] 
{Modelling turbulent fluxes due to thermal convection in rectilinear shearing flow}

\author[Radoslaw Smolec, G\"unter Houdek \& Douglas Gough]   
{Radoslaw Smolec$^1$, G\"unter Houdek$^1$
 \and Douglas Gough$^2$}

\affiliation{$^1$Institute of Astronomy, University of Vienna, A-1180 Vienna, Austria \\ 
email: {\tt radek.smolec@univie.ac.at, guenter.houdek@univie.ac.at} \\[\affilskip]
$^2$Institute of Astronomy and Department of Applied Mathematics and Theoretical Physics,\\ 
University of Cambridge, Cambridge CB3 0HA, UK \\email: {\tt douglas@ast.cam.ac.uk}}

\pubyear{2010}
\volume{271}  
\pagerange{1--2}
\jname{Astrophysical Dynamics: From Stars to Galaxies}
\editors{N. Brummell \& S. Brun, eds.}
\begin{document}

\maketitle

\vspace{-2pt}
\begin{abstract}
We revisit a phenomenological description of turbulent thermal convection
along the lines proposed originally by \cite[Gough (1965)]{DG65} in which 
eddies grow solely by extracting energy from the unstably stratified mean 
state and are subsequently destroyed by internal shear instability. This work
is part of an ongoing investigation for finding a procedure to calculate
the turbulent fluxes of heat and momentum in the presence of a shearing
background flow in stars. 

\end{abstract}
\noindent{\bf Introduction}

Convection models based on the mixing-length approach 
still represent the main method for computing the turbulent fluxes in stars 
with convectively unstable regions. In such regions the pulsational 
stability of the star is affected not only by the radiative heat flux but 
also by the modulation of the convective heat flux and by direct mechanical 
coupling of the pulsation with the convective motion via the Reynolds stresses. 
Time-dependent formulations of the mixing-length approach for radial 
pulsation have been proposed by, for example,  \cite[Gough (1965, 1977a)]{DG77a} 
and \cite[Unno (1967)]{U67}. In a first step towards a generalization to 
nonradially pulsating stars, \cite[Gough \& Houdek (2001)]{GH01} adopted 
Gough's formulation, incorporating into it a treatment 
of the influence of a shearing background flow. In this generalized framework 
of the mixing-length formalism, in which turbulent convective eddies grow 
according to linearized theory and are subsequently broken up by internal 
shear instability, there is a consequent reduction in the mean amplitude of 
the eddy motion, and a corresponding reduction in the heat flux.

In order to test and calibrate the formalism, it
is preferable first to compare its predictions with existing results of
hopefully more reliable investigations, such as experiments or numerical
simulations. Here we extend our earlier work \cite[(Gough \& Houdek 2001)]{GH01}, 
comparing the functional forms of our mean temperature and shear profiles 
with those of direct numerical simulations  
(DNS; \cite[Domaradzki \& Metcalfe 1988, DM88]{DM88}) of Rayleigh-B\'enard 
convection in air (Prandtl number, $\sigma=0.71$) with a strongly shearing 
background flow.  We consider a plane-parallel layer of fluid confined between 
rigid horizontal perfectly conducting boundaries of infinite extent at fixed 
temperatures, the lower being hotter than the upper by $\Delta T$. The upper 
boundary moves horizontally with constant velocity $\Delta U$, and we assume, 
in accordance with the Boussinesq approximation, that the shear, $E$, in the 
mean (plane Couette) flow does not vary over the scale of an eddy.

\medskip
\noindent{\bf Turbulent fluxes and mean equations in the presence of a shear} 

We follow the basic procedure by \cite[Gough \& Houdek (2001)]{GH01} to solve the linearized equations describing the dynamics in a statistically stationary flow of a viscous Boussinesq fluid confined between two horizontal planes. In Cartesian co-ordinates $(x,\,y,\,z)$ the equations are nondimensionalized using the vertical distance between the planes, $d$, and the thermal diffusion time across $d$ as units of space and time. Linearized modes of convection are obtained by expansion about $E:={\rm d}_zU=0$ (where ${\rm d}_z\equiv{\rm d}/{\rm d}z$) to second order in $E$. The resulting expressions for the eigenfunctions of the fluctuating temperature, $T'$, and turbulent velocity field, $\vec{u}=(u,v,w)$, are then used to compute the turbulent fluxes of heat, $\overline{wT'}$, and momentum (Reynolds stresses), $\overline{\rho u_iu_j}$, in the manner of \cite[Gough (1977a)]{DG77b}. A horizontal bar denotes statistical average (average over the horizontal plane).  
\begin{figure}[t]
\begin{center}
\includegraphics[width=5.1in]{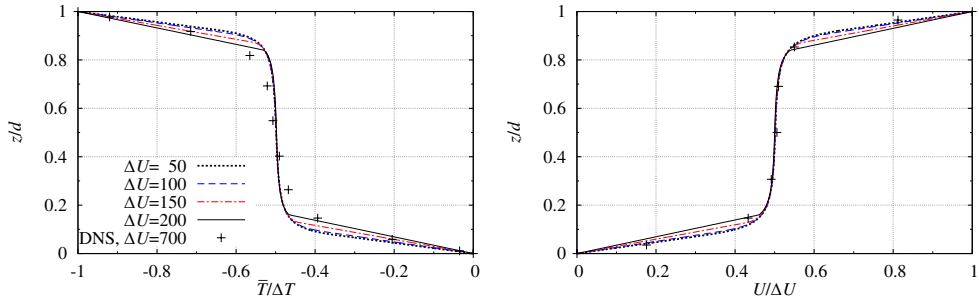} 
\caption{Mean temperature $\overline{T}/\Delta T$ (left panel) and mean velocity $U/\Delta U$ (right panel) as a function of height $z$ are compared with DNS data (crosses) for different $\Delta U$ values.}
\label{fig1}
\end{center}
\end{figure}
In our dimensionless formulation, the sum of the radiative and convective fluxes is independent of $z$ and equal to the Nusselt number of the layer, $N=F_{\rm r}(z)+F_{\rm c}(z)$. In the diffusion approximation the radiative flux is equal to the mean temperature gradient, $F_{\rm r}(z)=-{\rm d}_z\overline{T}$. The thermal mean equation is solved simultaneously with (the only non-trivial ($x$-) component of) the mean momentum equation: 
${\rm d}_z\overline{\rho uw}-\sigma{\rm d}_{zz}U=0$ (where ${\rm d}_{zz}\equiv{\rm d}^2/{\rm d}z^2$).
Note that the  Reynolds stress $\overline{\rho uw}$ distorts the shear, and consequently the $x$-component of the mean flow, $U$, is not a linear function of height $z$.

\medskip
\noindent{\bf Results and conclusions} 

Fig.~1 shows the normalized mean vertical velocity profiles, $U/\Delta U$ and mean temperature profiles, $\overline{T}/\Delta T$, for four values of $\Delta U$: 50, 100, 150 and 200. Our results are compared with DNS data (DM88), which assume $\Delta U=700$. The mean profiles are in reasonable agreement with the DNS data, but for smaller values of $\Delta U$ in our model computations. Best agreement with the DNS data is obtained with $\Delta U=150$. Two factors may be responsible for the discrepancy in the values of $\Delta U$: ($i$) to maintain greatest simplicity, we adopted horizontal-stress-free boundary conditions 
for the eddies, whereas in the simulations the bounding planes were taken to be rigid --  this may perhaps account for up to a factor of about three in the $\Delta U$ differences between the model ($\Delta U\!\!=\!150$) and the simulation ($\Delta U\!\!=\!700$); ($ii$) nonlocal effects may also contribute to the remaining differences between our results and the DNS data. We plan to investigate these issues, extending our model in the manner of \cite[Gough (1977b)]{DG77b} to accommodate nonlocal behaviour.

\medskip
\noindent
{\it This research is supported by the Austrian FWF grant No. AP 2120521. DOG 
is grateful to the Leverhulme Foundation for an Emeritus Fellowship.}

\end{document}